\documentclass[%
reprint,
aps,
superscriptaddress,
nofootinbib,
nobibnotes,
]{revtex4-1}

\usepackage{xcolor}
\usepackage{graphicx}
\graphicspath{{./fig/}{./fig/st_dgm/}{./fig/rela_st/}}
\usepackage{amsmath,amssymb,amsfonts,mathrsfs,bm}
\usepackage[%
colorlinks=true,
linkcolor=blue,
citecolor=blue,
]{hyperref}

\newcommand{\ee}{\mathrm{e}}
\newcommand{\ii}{\mathrm{i}}
\newcommand{\lie}{\mathscr{L}}
\newcommand{\br}{\bar}
\newcommand{\ck}{\check}
\newcommand{\tld}{\widetilde}
\newcommand{\ord}[2]{{}^{\mbox{\tiny$(\!#1\!)\!\!$}}{#2}}
\newcommand{\tl}[2]{{}^{\mbox{\tiny$#1\!$}}{#2}}

\newcommand{\Mp}{\mathcal{M}}
\newcommand{\Mr}{\br{\Mp}}
\newcommand{\Mb}{\mathcal{M}_{0}}
\newcommand{\Sp}{\mathcal{S}}
\newcommand{\Sr}{\br{\Sp}}

\newcommand{\Vp}{\mathcal{V}}

\newcommand{\GW}{\mathrm{GW}}

\newcommand{\cd}{\nabla}
\newcommand{\cds}{\mathcal{D}}
\newcommand{\cdr}{\br{\nabla}}
\newcommand{\cdvr}{\br{D}}

\newcommand{\gr}{\br{g}}
\newcommand{\gb}{\ord{0}{g}}
\newcommand{\tck}{\ck{t}}
\newcommand{\uck}{\ck{u}}
\newcommand{\vck}{\ck{v}}
\newcommand{\tr}{\br{t}}
\newcommand{\ur}{\br{u}}
\newcommand{\vrf}{\br{v}}

\begin{document}

\title{On the energy of gravitational waves}

\author{Rong-Gen Cai}
\affiliation{CAS Key Laboratory of Theoretical Physics, Institute of Theoretical Physics, Chinese Academy of Sciences, Beijing 100190, China}
\affiliation{School of Physical Sciences, University of Chinese Academy of Sciences, Beijing 100049, China}
\affiliation{School of Fundamental Physics and Mathematical Sciences, Hangzhou Institute for Advanced Study, University of Chinese Academy of Sciences, Hangzhou 310024, China}

\author{Xing-Yu Yang}
\email[Corresponding author.~]{yangxingyu@itp.ac.cn}
\affiliation{CAS Key Laboratory of Theoretical Physics, Institute of Theoretical Physics, Chinese Academy of Sciences, Beijing 100190, China}
\affiliation{School of Physical Sciences, University of Chinese Academy of Sciences, Beijing 100049, China}

\author{Long Zhao}
\affiliation{CAS Key Laboratory of Theoretical Physics, Institute of Theoretical Physics, Chinese Academy of Sciences, Beijing 100190, China}
\affiliation{School of Physical Sciences, University of Chinese Academy of Sciences, Beijing 100049, China}

\begin{abstract}
    The energy of gravitational waves is a fundamental problem in gravity theory.
    The existing descriptions for the energy of gravitational waves, such as the well-known Isaacson energy-momentum tensor, suffer from several defects.
    Due to the equivalence principle, the gravitational energy-momentum can only be defined quasilocally, being associated with a closed spacelike 2-surface bounding a region.
    We propose a new approach to derive the energy of gravitational waves \emph{directly} from the quasilocal gravitational energy.
    Such an approach is natural and consistent with the quasilocality of gravitational energy-momentum.
\end{abstract}

\maketitle

\section{Introduction}

One of great predictions of general relativity is the existence of gravitational waves (GWs).
The observation of GWs from a binary black hole merger~\cite{Abbott:2016blz} has launched a new era of astronomy and cosmology.
When discussing GWs, a fundamental problem is their energy.
Back to the 1950s, there was a controversy on whether or not GWs can carry energy.
The controversy was ultimately resolved by Bondi using a simple thought experiment~\cite{Bondi:1957dt}.

A mathematical description for the energy of GWs was not devised until the works of Isaacson, in which an effective energy-momentum tensor of GWs was obtained by averaging the square of gradient of the wave field over several wavelengths with the shortwave approximation~\cite{Isaacson:1967zz,Isaacson:1968zza}.
In applications to physics of the very early Universe, the fluctuations of interest have wavelengths larger than the Hubble radius, an effective energy-momentum tensor was derived by Mukhanov, Abramo and Brandenberger~\cite{Mukhanov:1996ak,Abramo:1997hu}.
In these approaches named as geometric approach, the gravitational field is divided into the background part and wave part, the effective energy-momentum tensor comes from the back reaction of wave to the background.
The other approach is named as field-theoretical approach, in which the effective energy-momentum tensor is derived by the Lagrange-Belinfante-Rosenfeld procedure~\cite{1939Phy.....6..887B,1940Phy.....7..449B,Rosenfeld.MARB18.1940.}.
The results are various expressions of pseudotensors~\cite{1915SPAW.......844E,Papapetrou:1948jw,Bergmann:1953jz,Goldberg:1958zz,1958AnPhy...4..347M,Landau:1982dva,Rosen:1993ea,Virbhadra:1995vu}.

Although different approaches for obtaining the energy of GWs were proposed in literature, they have several defects.
In the geometric approach, an artificial partition of the gravitational field is needed, while in the field-theoretical approach, the pseudotensor depends on coordinates.
Besides, an additional elaborate averaging scheme is necessary for both approaches in order to obtain a meaningful effective energy-momentum tensor of GWs.
The dependence on these artificial objects leads to some ambiguities.
As a result, different approaches are matched only in the linear order of metric perturbations.
These difficulties root in the equivalence principle of general relativity, which leads to the fact that the gravitational energy-momentum can not be defined locally.

The modern concept, introduced by Penrose~\cite{Penrose:1982wp}, is that a proper energy-momentum of gravity is quasilocal, being associated with a closed spacelike 2-surface bounding a region.
Chang, Nester and Chen found a natural quasilocal Hamiltonian interpretation of the pseudotensor~\cite{Chang:1998wj}.
Many proposals of quasilocal energy were made~\cite{1990CQGra...7..787K,Jezierski:1990vu,Dougan:1991zz,1992CQGra...9.1917B,Brown:1992br,Hayward:1993ph,Chen:1994qg,Chen:1998aw,Chen:2000xw,Liu:2003bx,Wang:2008jy}, of which an important one is the work of Brown and York in which they proposed their definition by using the Hamiltonian formulation of general relativity~\cite{Brown:1992br}.
The Brown-York energy has the right asymptotic behavior but is not nonnegative in general, this defect motivated the seminal works~\cite{Wang:2008jy,zbMATH05607355}, in which Wang and Yau proposed a well-defined quasilocal energy (for a comprehensive review see~\cite{Szabados:2009eka}).

GWs are just parts of the gravitational field itself, their energy should also be quasilocal.
A natural thought is that one should derive the energy of GWs \emph{directly} from the quasilocal gravitational energy instead of the ill-defined gravitational energy-momentum tensor.
To achieve this natural thought, one needs to deal with two questions:
(1) what is the proper description of gravitational energy?
(2) which parts of gravitational field can be taken as GWs?
In this paper, we deal with these two questions and give a new approach to derive the energy of GWs directly from the quasilocal gravitational energy.
Such an approach is more natural and more consistent with the quasilocality of gravitational energy-momentum.

\section{Quasilocal energy}

Due to the equivalence principle, the gravitational energy can not be defined locally.
However when there is an asymptotic symmetry, the total energy can be defined, which is called ADM energy and Bondi energy when viewed from spatial infinity and null infinity, respectively~\cite{Arnowitt:1961zz,Bondi:1962px}.
It was proposed to measure the energy of a system by enclosing it with a closed spacelike 2-surface, which is the idea behind the definition of quasilocal energy of the surface~\cite{Penrose:1982wp}.
There are several conditions the quasilocal energy should satisfy:
the ADM or Bondi energy should be recovered as spatial or null infinity is approached (`correct large-sphere behavior'),
the expected limits should be obtained when the surface converges to a point (`correct small-sphere behavior'),
the quasilocal energy should be nonnegative in general and vanish for Minkowski spacetime.
Brown and York obtained the quasilocal energy of a closed spacelike 2-surface including contributions from both gravitational and matter fields by employing a Hamilton-Jacobi analysis of the action functional~\cite{Brown:1992br}.
Their definition has the right asymptotic behavior and was proved to be positive when the spacelike 3-region which the surface bounds is time symmetric~\cite{zbMATH02171911}, but it depends on the choice of the 3-region and is not nonnegative in general.
Motivated by geometric consideration, Liu and Yau introduced a definition which is independent of the 3-region, and proved that it is always positive~\cite{Liu:2003bx}, however it can be strictly positive even when the surface is in a flat spacetime as pointed out by Murchadha, Szabados and Tod~\cite{2004PhRvL..92y9001M}.
After that, Wang and Yau rectified that defect and obtained a well-defined quasilocal energy~\cite{Wang:2008jy,zbMATH05607355}, which is the most satisfactory one so far.
Therefore, for the question (1) what is the proper description of gravitational energy, we adopt the quasilocal energy given by Wang and Yau, and we will briefly revisit it here.

For a closed spacelike 2-surface $\Sp$ which bounds a spacelike 3-region $\Vp$ in the physical spacetime $(\Mp, g_{ab})$.
Let $u^{ a }$ be the future-directed timelike unit normal to $\Vp$ and $v^{ a }$ be the outward spacelike unit normal to $\Sp$ such that $u^{ a }v_{ a }=0$.
Given a future-directed timelike unit vector field $t^{ a }$ along $\Sp$, it can be decomposed as $t^{ a }=N u^{ a }+N^{ a }$.
Through the Hamilton-Jacobi analysis of the gravitational action, one can obtain the surface Hamiltonian
\begin{equation}
    \mathfrak{H}(t^{ a }, u^{ a }) = -\frac{1}{\kappa} \int_{\Sp} [Nk - N^{ a }v^{ b }(K_{ ab }-K^{ c }_{ c } \gamma_{ ab })],
\end{equation}
where $\kappa=8\pi G/c^{4}$, $k$ is the trace of the two-dimensional extrinsic curvature of $\Sp$ in $\Vp$ with respect to $v^{ a }$, $K_{ ab }$ is the extrinsic curvature of $\Vp$ in $\Mp$ with respect to $u^{ a }$, and $\gamma_{ ab }$ is the induced Riemannian metric on $\Vp$.

To define the quasilocal energy, one needs to find a reference action that corresponds to fixing the metric on the timelike boundary of the time history of the bounded region, and compute the corresponding reference surface Hamiltonian $\br{\mathfrak{H}}$.
The quasilocal energy is then the difference of the two surface Hamiltonian $E=\mathfrak{H}-\br{\mathfrak{H}}$.
The reference spacetime is arbitrary, and the freedom to choose different reference spacetime is just the freedom to choose the zero point of energy for the system.
Therefore considering the conditions that quasilocal energy should satisfy, a natural choice for the reference spacetime $(\Mr,\gr_{ ab })$ is Minkowski spacetime, which is used through the paper.%

Given only $\Sp$ and $t^{ a }$ in the physical spacetime $(\Mp, g_{ ab })$, in order to calculate the quasilocal energy of $\Sp$ with respect to the observer $t^{ a }$, Wang and Yau propose the following procedures.

Viewing $t^{ a }$ as a vector field along $\Sp$, one can decompose $t^{ a }$ as $t^{ a }=\tl{\perp}{t}^{ a } + \tl{\parallel}{t}^{ a }$, where $\tl{\parallel}{t}^{ a }$ is tangent to $\Sp$ and $\tl{\perp}{t}^{ a }$ is normal to $\Sp$.
The tangent vector $\tl{\parallel}{t}^{ a }$ can be written as $\tl{\parallel}{t}^{ a } = -\cds^{ a }\tau$ where $\cds$ is covariant derivative of the induced Riemannian metric $\sigma_{ ab }$ on $\Sp$.
In \cite{zbMATH05607355}, Wang and Yau proved the following existence theorem.
Given a Riemannian metric $\sigma_{ ab }$ and a function $\tau$ on $\Sp$ (which is topologically $S^{2}$) such that the metric $\sigma+ d\tau \otimes d\tau$ is a Riemannian metric with positive Gaussian curvature.
There exists a unique spacelike isometric embedding $\varphi:\Sp \rightarrow \Mr$,
 i.e. $\sigma_{ ab }=(\varphi^{*}\br{\sigma})_{ ab }$ where $\br{\sigma}_{ ab }$ is the induced Riemannian metric on $\Sr \equiv \varphi(\Sp)$,
such that on $\Sr$ the time function (with respect to a future-directed timelike unit translational Killing vector field $\tr^{ a }$ in $\Mr$) restricts to $\tau$.

Viewing $\tr^{ a }$ as a vector field along $\Sr$, one can decompose $\tr^{ a }$ as $\tr^{ a } = \tl{\perp}{\tr}^{ a } + \tl{\parallel}{\tr}^{ a }$, where $\tl{\parallel}{\tr}^{ a }$ is tangent to $\Sr$ and $\tl{\perp}{\tr}^{ a }$ is normal to $\Sr$.
Choose $\ur^{ a }= \tl{\perp}{\tr}^{ a }/|\tl{\perp}{\tr}^{ a }|$, $\br{N}=|\tl{\perp}{\tr}^{ a }|$ and $\vrf^{ a }$ be the outward spacelike unit normal vector of $\Sr$ that is orthogonal to $\ur^{ a }$, then one can obtain the corresponding surface Hamiltonian $\br{\mathfrak{H}}(\tr^{ a },\ur^{ a })$.

Having picked $\tr^{ a }$ and $\ur^{ a }$ along $\Sr$ in $\Mr$, one chooses a corresponding pair of $\tck^{ a }$ and $\uck^{ a }$ along $\Sp$ in $\Mp$ as follows.
First, via the embedding $\varphi$, $\tl{\parallel}{\tr}^{ a }$ can be pulled back to $\Sp$ and be viewed as a vector field tangent to $\Sp$ in $\Mp$.
Next, define  $\tck^{ a } = \ck{N} \uck^{ a } + \tl{\parallel}{\tck}^{ a }$, where $\ck{N} \equiv \varphi^{*} \br{N}$ ,$\tl{\parallel}{\tck}^{ a } \equiv (\varphi^{*} \tl{\parallel}{\tr})^{ a }$, and $\uck^{ a }$ is the unique future-directed timelike unit normal vector along $\Sp$ such that $\mathfrak{h}_{ a } \uck^{ a } = \br{\mathfrak{h}}_{ a } \ur^{ a }$.
Here $\mathfrak{h}_{ a }$ is the mean curvature vector of $\Sp$ in $\Mp$ which is assumed to be spacelike, and $\br{\mathfrak{h}}^{ a }$ is the mean curvature vector of $\Sr$ in $\Mr$.
The mean curvature vector can be calculated as $\mathfrak{h}^{ a }=(K^{ c }_{ c }-K_{de}v^{d}v^{e})u^{ a }-k v^{ a }$, which is indeed independent of the choice of $u^{ a }$ and $v^{ a }$.
Physically, $\mathfrak{h}_{ a } \uck^{ a } = \br{\mathfrak{h}}_{ a } \ur^{ a }$ means the expansions of $\Sp \subset \Mp$ and $\Sr \subset \Mr$ along the respective directions $\uck^{ a }$ and $\ur^{ a }$ are the same.
When the observer $t^{a}$ is orthogonal to $\Vp$, i.e. $t^{a}=u^{a}$, one has $\tck^{a}=\uck^{a}=u^{a}=t^{a}$, such an observer is also called Eulerian observer~\cite{2012LNP...846.....G}, therefore the difference between $u^{a}$ and $\uck^{a}$ shows the difference between Eulerian and non-Eulerian observers.
Take $\vck^{ a }$ to be the spacelike unit normal vector of $\Sp$ that is orthogonal to $\uck^{ a }$ and satisfies $\vck^{ a } \mathfrak{h}_{ a }<0$.
The quasilocal energy of $\Sp$ with respect to the observer $t^{ a }$ is
$E(\Sp,t^{ a }) = E(\Sp,\tau) = \mathfrak{H}(\tck^{ a },\uck^{ a }) - \br{\mathfrak{H}}(\tr^{ a },\ur^{ a })$,
which depends on $t^{ a }$ only through $\tau$.

We here summarize the physical essence briefly without any cumbersome mathematical description.
Due to the equivalence principle, the energy of a physical system should be measured by enclosing it with a closed spacelike 2-surface $\Sp$.
Note that energy is a conserved quantity, but not an invariant quantity, it depends on the observer.
Given $\Sp$ in physical spacetime $(\Mp, g_{ab})$, one can only obtain its energy with respect to certain observer $t^{a}$.
Having the closed spacelike 2-surface and the observer, the energy of the physical system with respect to the observer is determined, which can be calculated by the aforementioned mathematical procedure.

\section{Gravitational waves}

\begin{figure}[htpb]
    \centering
    \includegraphics{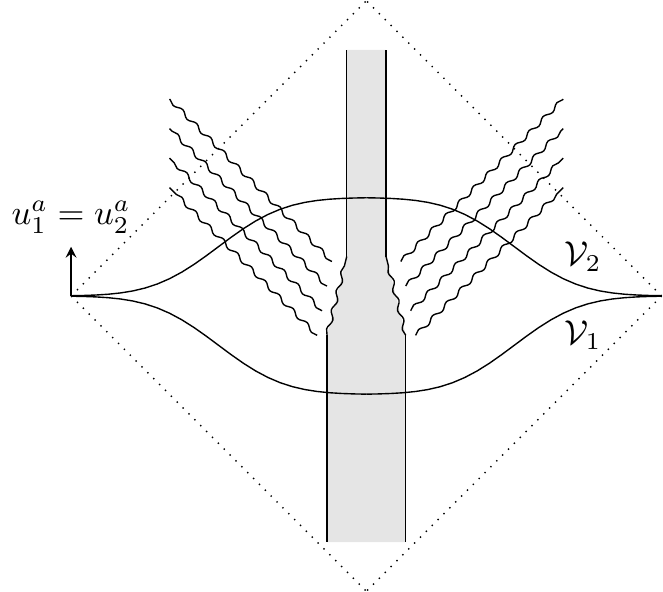}
    \caption[]{Schematic diagram for a pulse of GWs.
        The shaded part denotes the worldtube of the source and the wavy lines denote the GWs.
        $\Vp_{1}$ and $\Vp_{2}$ are two different spacelike 3-regions with same boundary and same normal vector on the boundary.
}
\label{fig:st_dgm}
\end{figure}

Although only the closed spacelike 2-surface $\Sp$ is involved when concerning the total energy, the spacelike 3-region $\Vp$ must be specified when GWs are concerned, which can be easily seen as follows.
Suppose a source emits a pulse of GWs, then for two different $\Vp_{1}$ and $\Vp_{2}$ with same boundary $\Sp$ as shown in Fig.~\ref{fig:st_dgm}, they have same total energy but different energy of GWs.

In order to derive the GW energy from gravitational energy, one needs to deal with the question (2) which parts of gravitational field can be taken as GWs.
In other words, one needs to specify the quantity that can be taken as GWs.
However, this is not as easy as it seems to be.
GWs in the most general sense are time-dependent solutions of the Einstein equations, but this definition of waves is rather too broad, for a field which changes only as a result of the relative motion of the source and the observer (motion past a static field) would not be called a wave.
Most additional demands which GWs should satisfy lead, however, to the characterization `radiation or transport of energy', and this is where the difficulties begin, starting with the definition of energy~\cite{Stephani:2004ud}.
Specifically, following questions arises naturally:
(a) for a point in a given spacetime, can one say whether there are GWs pass through this point?
(b) for a point with a observer in a given spacetime, can one say whether there are GWs pass through this point?
(c) for a spacelike 3-region in a give spacetime, can one say whether there are GWs in this 3-region?
If the answer of (a) is yes, then the GWs should only depend on the metric, which means that the GWs are intrinsic structure of the spacetime.
If the answer of (a) is no while the answer of (b) is yes, then the GWs should depend on the metric and the observer, which means that the GWs by definition are observer dependent.
If the answer of (a,b) is no while the answer of (c) is yes, then the GWs should depend on the metric and the spacelike 3-region, which means that the GWs can not be defined locally.

Pirani proposed a criterion for the existence of gravitational radiation: at any point in vacuum spacetime, gravitational radiation is present if the Weyl tensor is of Petrov type II or N or III, but not if it is of Petrov type I or D~\cite{Pirani:1956wr,1954UZKGU.114...55P}.
With this criterion, one can give a yes answer for the question (a) by the Petrov type of Weyl tensor for a point in a given spacetime.
However, since the radiation fields with realistic sources have the algebraically most general structure (Petrov type I), this criterion is too stringent~\cite{Sachs:1961zz}.

For a point with observer $t^{a}$ in the spacetime, the equation of geodesic deviation reads
\begin{equation}
    a^{a} = -R^{a}{}_{bcd} t^{b} t^{d} s^{c} = -C^{a}{}_{bcd} t^{b} t^{d} s^{c} + \text{Ricci terms},
\end{equation}
where $a^{a} \equiv t^{b}\nabla_{b}(t^{c}\nabla_{c}s^{a})$ and $C_{abcd}$ is the Weyl tensor.
Choosing an orthonormal tetrad $\{(e_{\mu})^{a}\}$ such that $(e_{0})^{a}=t^{a}$, one can construct a tetrad of four null basis vectors $\{m^{a},\br{m}^{a},l^{a},k^{a}\}$ by defining
$m^{a} \equiv [ (e_{1})^{a} - \ii (e_{2})^{a} ]/\sqrt{2}$,
$\br{m}^{a} \equiv [ (e_{1})^{a} + \ii (e_{2})^{a} ]/\sqrt{2}$,
$l^{a} \equiv [ (e_{0})^{a} - (e_{3})^{a} ]/\sqrt{2}$,
$k^{a} \equiv [ (e_{0})^{a} + (e_{3})^{a} ]/\sqrt{2}$~\cite{Newman:1961qr}(here and hereafter we follow the notations in~\cite{Stephani:2003tm}).
The tetrad components of $C_{abcd} t^{b} t^{d}$ is $(e_{\mu})^{a} (e_{\nu})^{c} C_{abcd} t^{b} t^{d}$, which can be written as
\begin{equation} \label{eq:tetrad_com_Ctt}
    \begin{pmatrix}
        0 & 0 & 0 & 0 \\
        0 & \Psi_{0}^{\Re}/2 + \Psi_{4}^{\Re}/2 - \Psi_{2}^{\Re} & -\Psi_{0}^{\Im}/2 + \Psi_{4}^{\Im}/2 & -\Psi_{1}^{\Re}+\Psi_{3}^{\Re} \\
        0 & -\Psi_{0}^{\Im}/2 + \Psi_{4}^{\Im}/2 & -\Psi_{0}^{\Re}/2-\Psi_{4}^{\Re}/2-\Psi_{2}^{\Re} & \Psi_{1}^{\Im}+\Psi_{3}^{\Im} \\
        0 & -\Psi_{1}^{\Re}+\Psi_{3}^{\Re} & \Psi_{1}^{\Im}+\Psi_{3}^{\Im} & 2 \Psi_{2}^{\Re}
    \end{pmatrix}
\end{equation}
where $\{\Psi_{0},\cdots,\Psi_{4}\}$ are the Weyl scalars and the superscripts $\Re$ and $\Im$ denote their real and imaginary parts respectively.
Eq.~\eqref{eq:tetrad_com_Ctt} shows that the Weyl scalars $\Psi_{2}$, $\{\Psi_{1},\Psi_{3}\}$, $\{ \Psi_{0},\Psi_{4} \}$ represent the Coulomb, longitudinal, transverse parts of gravitational field respectively, as pointed out by Szekeres~\cite{Szekeres:1965ux}.
Therefore $\{ \Psi_{0},\Psi_{4} \}$ are usually taken as GWs.
However, the Weyl scalars depend on the choice of tetrad.
Even though $(e_{0})^{a}$ is determined by observer $t^{a}$, one still has freedom to change $\{(e_{1})^{a}, (e_{2})^{a}, (e_{3})^{a}\}$.

Due to its algebraic structure, the Weyl tensor can be written as canonical form with certain tetrad.
Such a tetrad is determined up to a type iii tetrad transformation for Petrov type (I,II,III,D), and is determined up to a type ii or iii tetrad transformation for Petrov type N, therefore the null direction $k^{a}$ is uniquely determined (see Appendices)~\cite{Stephani:2003tm}.
This unique null direction represents the null direction of GWs, and coincides with one of the principal null directions for type (II,III,D,N) but does not for type I.
In other words, for all the nontrivial Petrov type (I,II,III,D,N), there is a \emph{unique} null direction $\mathbb{K}^{a}$ given by the Weyl tensor itself.
As to the Petrov type O, it is conformally flat due to the vanishing Weyl tensor and has no GWs.

Now, for any point with observer $t^{a}$ in the given spacetime, one can set a tetrad such that $(e_{0})^{a}=t^{a}$ and $k^{a} \equiv [ (e_{0})^{a} + (e_{3})^{a} ]/\sqrt{2}$ is in the same direction of $\mathbb{K}^{a}$, we call such a tetrad as \emph{canonical observer tetrad}.
There is only one degree of freedom left for such a tetrad, i.e. $m^{a} \rightarrow \ee^{\ii\theta}m^{a}$, thus the Weyl scalars are determined up to a pure phase such as $\Psi_{4} \rightarrow \ee^{-2\ii\theta}\Psi_{4}$.

The physical meaning of the canonical observer tetrad is related to the properties of GWs, and can be understood as follows.
For an observer $t^{a}$ in the Petrov type N pp-waves spacetime, the GWs with respect to this observer are physically determined.
However, if arbitrary tetrad is used, nonvanishing $\{ \Psi_{2},\Psi_{1},\Psi_{3} \}$ could be obtained, this does not mean there are Coulomb and longitudinal gravitational field in this spacetime, it is just a result of the improper choice of tetrad.
Once the canonical observer tetrad is used, the only nonvanishing Weyl scalar is $\Psi_{4}$ which represents the pp-waves.
Therefore, \emph{not any} Weyl scalars $\{ \Psi_{0},\Psi_{4} \}$ can be taken as GWs, \emph{only} the Weyl scalars $\{ \Psi_{0},\Psi_{4} \}$ in the canonical observer tetrad can be physically taken as GWs.

For a spacelike 3-region $\Vp$ with closed boundary $\Sp$ in given spacetime, the induced Riemannian metric $\gamma_{ab}$ on $\Vp$ is determined.
Choosing a 3-dimensional Riemannian metric $\tld{\gamma}_{ab}$ on $\Vp$ which satisfies $R^{(\tld{\gamma})}_{ab}=c \tld{\gamma}_{ab}$ for some constant $c$, then $(\Vp, \tld{\gamma}_{ab})$ is a compact Riemannian Einstein space, and $\gamma_{ab}$ is just a second rank symmetric tensor field on $(\Vp, \tld{\gamma}_{ab})$.
It is proved that any second rank symmetric tensor field on $(\Vp, \tld{\gamma}_{ab})$, of course including $\gamma_{ab}$, can be uniquely decomposed as~\cite{Ishibashi:2004wx}
\begin{equation}
    \gamma_{ab} = H_{ab} + \tld{D}_{(a} F_{b)} + \left( \tld{D}_{a}\tld{D}_{b} -\frac{1}{3} \tld{\gamma}_{ab} \tld{D}^{c}\tld{D}_{c} \right)E + C \tld{\gamma}_{ab} ,
\end{equation}
where $\tld{D}^{a}H_{ab}=0=H^{a}{}_{a}$, $\tld{D}^{a}F_{a}=0$, $\tld{D}$ is the covariant derivative of $\tld{\gamma}_{ab}$, and the indices are raised and lowered with $\tld{\gamma}_{ab}$.
$H_{ab}$, $F_{a}$ and $\{E,C\}$ are tensor-, vector- and scalar-type components of $\gamma_{ab}$ respectively.
The tensor-type component $H_{ab}$ of $\gamma_{ab}$ is usually taken as GWs.
However, $H_{ab}$ depends on the choice of $\tld{\gamma}_{ab}$.
Here we choose a 3-dimensional flat metric for $\tld{\gamma}_{ab}$, there is still freedom of diffeomorphism $\phi:\Vp \rightarrow \Vp$, namely, $(\Vp,\gamma_{ab},\tld{\gamma}_{ab})$ and $(\Vp,\gamma_{ab},(\phi^{*}\tld{\gamma})_{ab})$ could give different $H_{ab}$.

Recalling the unique null direction $\mathbb{K}^{a}$ given by the Weyl tensor, the projection of this null direction onto $\Vp$ gives a unique spacelike unit tangent field $\bm{k}^{a}$ on $\Vp$.
The field $\bm{k}^{a}$ is intrinsic on $\Vp$, and represents the spatial direction of GWs, which gives a requirement $\bm{k}^{a}H_{ab}=0$.
Such a requirement determines $\tld{\gamma}_{ab}$ up to a rotation around $\bm{k}^{a}$, and the one left degree of freedom is similar to that of tetrad $m^{a} \rightarrow \ee^{\ii\theta}m^{a}$, we call such a $\tld{\gamma}_{ab}$ as \emph{canonical 3-metric}.
The physical meaning of the canonical 3-metric is similar to that of canonical observer tetrad, it means that \emph{not any} $H_{ab}$ can be taken as GWs, \emph{only} the $H_{ab}$ in the canonical 3-metric can be physically taken as GWs.

In summary, when talking about GWs, one has to specify the quantity that can be taken as GWs.
In principle, as a physical quantity, it should be a scalar or vector or tensor on the spacetime, and should only depend on the metric and observer.
As is shown above, in a given spacetime, for a point with observer, the Weyl scalars $\{ \Psi_{0},\Psi_{4} \}$ in the canonical observer tetrad can be taken as GWs, for a spacelike 3-region $\Vp$ with closed boundary $\Sp$, the tensor component $H_{ab}$ in canonical 3-metric can be taken as GWs.
\emph{One should understand the canonical observer tetrad and canonical 3-metric as particular procedures to construct the quantities that can be physically taken as GWs.}
The Weyl scalars describe GWs based on curvature while the tensor component describes GWs based on metric, these two descriptions are known to be consistent in the linearized theory~\cite{Bishop:2016lgv}, and the consistency beyond linearized theory needs further studies.
Although there is one degree of freedom left, it seems that this left degree of freedom can not be removed due to the transverseness of GWs, fortunately it does not affect the GW energy as will be shown in the next section.

\section{Energy of gravitational waves}

GWs are just parts of the gravitational field, a natural thought is that one should derive the energy of GWs directly from the gravitational energy.
To achieve this natural thought, one needs to deal with two questions:
(1) what is the proper description of gravitational energy?
(2) which parts of gravitational field can be taken as GWs?
For the first question, we adopt the quasilocal energy given by Wang and Yau.
For the second question, we have shown that,
in a given spacetime, for a point with observer, the Weyl scalars $\{ \Psi_{0},\Psi_{4} \}$ in the canonical observer tetrad can be taken as GWs, for a spacelike 3-region $\Vp$ with closed boundary $\Sp$, the tensor component $H_{ab}$ in canonical 3-metric can be taken as GWs.

In the small surface limit, the quasilocal gravitational energy $E(\Sp,t^{a}) \propto r^{3} T_{ab}t^{a}t^{b} + \mathcal{O}(r^{4})$ for non-vacuum spacetime and $E(\Sp,t^{a}) \propto r^{5} T_{abcd}t^{a}t^{b}t^{c}t^{d} + \mathcal{O}(r^{6})$ for vacuum spacetime, where $T_{ab}$ is the energy-momentum tensor of matter fields and $T_{abcd}$ is the Bel-Robinson tensor~\cite{Szabados:2009eka,Chen:2015par}.
This manifestly shows the nonlocality of the gravitational energy.
The Bel-Robinson tensor
\begin{equation}
    T_{abcd} = C_{a}{}^{e}{}_{c}{}^{f} C_{bedf} + C_{a}{}^{e}{}_{d}{}^{f} C_{becf} - \frac{1}{2} g_{ab} C_{c}{}^{efg} C_{defg}
\end{equation}
has similar properties to the energy-momentum tensor, and sometimes is called superenergy tensor~\cite{Senovilla:1999xz}.
In the canonical observer tetrad, one has
\begin{equation}
    T_{abcd}t^{a}t^{b}t^{c}t^{d} = \Psi_{0}\br{\Psi}_{0} + 4\Psi_{1}\br{\Psi}_{1} + 6\Psi_{2}\br{\Psi}_{2} + 4\Psi_{3}\br{\Psi}_{3} + \Psi_{4}\br{\Psi}_{4}.
\end{equation}
Therefore, the energy of GWs in the small sphere limit is
\begin{equation}
E_{\GW}(t^{a}) \propto r^{5}(\Psi_{0}\br{\Psi}_{0} + \Psi_{4}\br{\Psi}_{4}).
\end{equation}
It is obvious that the one left degree of freedom for the canonical observer tetrad does not affect the GW energy, since it contributes a pure phase to the Weyl scalars.

For a spacelike 3-region $\Vp$ with closed boundary $\Sp$, the quasilocal gravitational energy is $E(\Sp,t^{a}) = \mathfrak{H} - \br{\mathfrak{H}}$.
The surface Hamiltonian $\mathfrak{H}$ is calculated in the physical spacetime which contains the information of GWs, while the surface Hamiltonian $\br{\mathfrak{H}}$ is calculated in the reference spacetime which represents the zero point of energy.

Recalling the isometric embedding $\varphi:\Sp \rightarrow \Mr$, one can extend it into a mapping $\Phi:\Mp \rightarrow \Mr$ such that $g_{ab}=(\Phi^{*}\br{g})_{ab}$ on $\Sp$ and $\tck^{a}=(\Phi^{*}\tr)^{a}$ on $\Vp$ by exponential map~\cite{Liu:2017neh}, where $\Phi^{*}$ is the pullback and will be omitted for brevity when there is no confusion.
With the mapping $\Phi$, one can pullback the tensors on reference spacetime to the physical spacetime, then the calculations of $E=\mathfrak{H} - \br{\mathfrak{H}}$ can be performed only on the physical spacetime.

It is found that the quasilocal energy derived by Wang and Yau is closely related to that derived by Chen, Nester and Tung from a covariant Hamiltonian formalism~\cite{Chen:1994qg,Chen:2005hwa,Liu:2017neh}, which is closely related to the conserved currents obtained by Katz, {Bi{\v{c}}{\'a}k} and {Lynden-Bell} by applying the Noether theorem to the Lagrangian~\cite{1997PhRvD..55.5957K,Chen:2000xw}.
Inspired by such relations, one can obtain
\begin{equation}\label{eq:WY-KBLB}
    E(\Sp,t^{a}) = \int_{\Sp} J^{ ab } \vck_{ b } \uck_{ a } = \int_{\Vp} I^{ a } \uck_{ a }
\end{equation}
where $I^{ a }=\cd_{ b } J^{ ab }$ is the conserved Noether current which satisfies $\cd_{ a } I^{ a } \equiv 0$.
In Eq.~\eqref{eq:WY-KBLB}, the first equality comes from the relation of quasilocal energy given by Wang-Yau and Katz-{Bi{\v{c}}{\'a}k}-{Lynden-Bell}, while the second equality comes from the Stokes' theorem.

Furthermore, $I^{ a }$ can be written as $I^{ a } = \left( T^{ a }_{ b } + \mathcal{T}^{ a }_{ b } \right) \tck^{ b }$, where $T^{ a }_{ b }$ is the energy-momentum tensor of matter fields which satisfies the Einstein equation $G^{ a }_{ b }=\kappa T^{ a }_{ b }$, and
\begin{equation}
    \begin{aligned}
        2\kappa\mathcal{T}^{ a }_{ b } = g^{ de } [ ( \Delta^{ c }_{ d  c }\Delta^{ a }_{ e  b } + \Delta^{ a }_{ de }\Delta^{ c }_{ c  b } - 2\Delta^{ a }_{ d  c }\Delta^{ c }_{ e  b } ) \\
        - \delta^{ a }_{ b } ( \Delta^{ f }_{ de }\Delta^{ c }_{ f  c } - \Delta^{ f }_{ d  c }\Delta^{ c }_{ f  e } ) ] \\
        + g^{ a  c } ( \Delta^{ e }_{ de }\Delta^{ d }_{ c  b } - \Delta^{ e }_{ c  e }\Delta^{ d }_{ d  b } ),
    \end{aligned}
\end{equation}
with
$
    \Delta^{ c }_{ ab }
    =\frac{1}{2} g^{ c  d } \left( \cdr_{ a }g_{ d  b } + \cdr_{ b }g_{ d  a } - \cdr_{ d }g_{ ab } \right)
$
is the difference between the Christoffel symbols in $\Mp$ and $\Mr$~\cite{1997PhRvD..55.5957K}.
One can find that the energy of a physical system contains contributions from matter fields and gravitational fields, and the energy of GWs shall contribute the part of the gravitational energy.

The energy of GWs in $\Vp$ with respect to observer $t^{a}$ is the part, which is contributed only by the tensor component $H_{ ab }$, of the gravitational energy,
\begin{equation}\label{eq:energy_GWs}
    E_{\GW}(\Vp,t^{a}) = \int_{\Vp} \mathfrak{P}_{H} \mathcal{T}^{ a }_{ b } \tck^{ b } \uck_{ a },
\end{equation}
where the operator $\mathfrak{P}_{H}$ picks out the parts of $\mathcal{T}^{ a }_{ b }$ that only depend on $H_{ ab }$.

On the 3-region $\Vp$ in the physical spacetime, there are two kinds of 3-metric $\{\gamma_{ab}, \tld{\gamma}_{ab} \}$ and $(\Phi^{*}\br{\gamma})_{ab}$.
The physical 3-metric $\gamma_{ab}$ contains the physical information and the canonical 3-metric $\tld{\gamma}_{ab}$ is related to the properties of GWs, the combination of these two 3-metric gives the quantity that can be taken as GWs, i.e. $\{\gamma_{ab}, \tld{\gamma}_{ab} \} \Rightarrow H_{ab}$.
On the other hand, in principle, $\Phi$ has nothing to do with the GWs, so does $(\Phi^{*}\br{\gamma})_{ab}$, it is only related to the gravitational energy.
However, one can always choose a $\Phi$ such that $(\Phi^{*}\br{\gamma})_{ab} = \tld{\gamma}_{ab}$, then $\Phi$ is not only related to gravitational energy but also related to GWs, we call such a $\Phi$ as \emph{canonical mapping}.
Within the canonical mapping, the calculation of Eq.~\eqref{eq:energy_GWs} can be greatly simplified.
One may think that the existence of the canonical mapping contradicts the general covariance, but it does not.
\emph{The canonical mapping should be understood as a particular procedure to construct a tensor field $\br{\gamma}_{ab}$ on $\Vp$.}
After the tensor field $\br{\gamma}_{ab}$ is obtained, there are two tensor fields $\{\gamma_{ab},\br{\gamma}_{ab}\}$ on $\Vp$, one can use them to discuss the gravitational energy and GWs.
The general covariance is preserved obviously since $\{\gamma_{ab},\br{\gamma}_{ab}\}$ are just two ordinary tensor fields on the manifold $\Mp$, and the diffeomorphism $\mathcal{X}:\Mp \rightarrow \Mp$ just pullback $\{\gamma_{ab},\br{\gamma}_{ab}\}$ to $\{(\mathcal{X}^{*}\gamma)_{ab},(\mathcal{X}^{*}\br{\gamma})_{ab}\}$.

Note that $\mathfrak{P}_{H} \mathcal{T}^{ a }_{ b }$ can be taken as an effective energy-momentum tensor of GWs, but it depends on the observer unlike the energy-momentum tensor of matter fields.
In our approach, the energy of GWs is derived \emph{directly} from the quasilocal gravitational energy, then one can call a tensor as effective energy-momentum tensor if the volume integration of its time-time component equals to the energy.
This is a significant difference between our approach and the existing approaches in literature, in which one derives an effective energy-momentum tensor (which suffers from several defects) firstly, then calls the volume integration of its time-time component as energy.

\section{Calculations in perturbation theory}

Usually in practice, it is hard to get the physical metric $g_{ab}$ by solving the Einstein equation directly.
Regarding that the physical metric deviates smally from a known exact solution which is the so-called background spacetime $(\Mb, \gb_{ab})$, one can apply the perturbation theory to find the approximate solution of the Einstein equation.
In order to make the comparison of tensors in the physical and background spacetime meaningful, a prescription for identifying points of these spacetimes must be given.
A gauge choice is precisely this, i.e., a mapping $\Upsilon$ between the physical and background spacetime.
Using a gauge $\Upsilon$, the physical metric $g_{ab}$ can be perturbed on the background metric $\gb_{ab}$ as $g_{ab} = \sum_{n=0}^{\infty} \frac{1}{n!} \ord{n}{g}_{ab}$, then the energy of GWs can be calculated order by order, $E_{\GW}=\sum_{n=0}^{\infty}\ord{n}{E}_{\GW}=\sum_{n=0}^{\infty} \int_{\Vp} \mathfrak{P}_{H} \ord{n}{\mathcal{T}^{a}_{b}} \tck^{b} \uck_{a}$.

Here we show an explicit calculation for GWs in asymptotically flat spacetime with a vacuum Einstein equation.
For such a spacetime, a convenient choice for the background spacetime is Minkowski spacetime.
Consider a 3-region $\Vp$ with Eulerian observer $t^{a}$, calculating $\mathfrak{P}_{H} \mathcal{T}^{a}_{b}$ order by order gives $\mathfrak{P}_{H} \ord{0}{\mathcal{T}}^{a}_{b}=\mathfrak{P}_{H} \ord{1}{\mathcal{T}}^{a}_{b}=0$, and
\begin{equation}
    \begin{aligned}
        2\kappa \mathfrak{P}_{H} \ord{2}{\mathcal{T}}^{a}_{b} t^{b} t_{a} &= \frac{1}{4}\ord{1}{\dot{H}}_{ab}\ord{1}{\dot{H}}^{ab} + \frac{1}{4}\cdvr_{c}\ord{1}{H}_{ab}\cdvr^{c}\ord{1}{H}^{ab} \\
                                                 &- \frac{1}{2}\cdvr_{a}\ord{1}{H}_{cb}\cdvr^{b}\ord{1}{H}^{ca},
    \end{aligned}
\end{equation}
where the over dot denotes the Lie derivative $\lie_{\tr}$.
The first and second terms are equal due to the first order Einstein equation.
The volume integration of the last term can be converted to a surface integral after integration by parts, which vanishes due to the isometric condition $g_{ab}=\gr_{ab}$ on the surface.
Then one has
\begin{equation}\label{eq:EGW2}
    \ord{2}{E}_{\GW}(\Vp,t^{a}) = \int_{\Vp} \frac{1}{4\kappa} \ord{1}{\dot{H}}_{ab}\ord{1}{\dot{H}}^{ab},
\end{equation}
which is the commonly used form in literature.

As explained in the last section, after the constructing procedure given by the canonical mapping, one has two tensor fields $\{\gamma_{ab},\br{\gamma}_{ab}\}$ on $\Vp$.
The energy of GWs is completely determined by these two tensor fields on $\Vp$ and has nothing to do with the background spacetime $(\Mb, \gb_{ab})$ and the gauge $\Upsilon$.
The perturbation theory is just auxiliary mathematical tool for calculations and is not necessary in principle, therefore it is obvious that the GW energy is independent of perturbation theory and is gauge invariant.
In a general case, the choice of background spacetime is quite arbitrary as long as the perturbation theory is valid.
For instance, the FLRW metric is commonly used as the background metric in the cosmological perturbation theory.

\section{Conclusions}

The energy of GWs is a fundamental problem in gravity theory.
The existing descriptions for the energy of GWs, such as the well-known Isaacson energy-momentum tensor, suffer from several defects.
Due to the equivalence principle, the gravitational energy-momentum can not be defined locally in general relativity.
The proper gravitational energy-momentum is quasilocal, being associated with a closed spacelike 2-surface bounding a region, as introduced by Penrose.

GWs are just parts of the gravitational field, a natural thought is that one should derive the energy of GWs \emph{directly} from the gravitational energy.
To achieve this natural thought, we deal with two questions:
(1) what is the proper description of gravitational energy?
(2) which parts of gravitational field can be taken as GWs?
For the first question, we adopt the quasilocal energy given by Wang and Yau, for the second question we specify the quantity that can be taken as GWs by introducing a constructing procedure.
With these two questions settled, we derive the energy of GWs directly from the quasilocal gravitational energy.
Such a quasilocal approach is more natural and more consistent with the quasilocality of gravitational energy-momentum.

Although we only show an explicit calculation for vacuum asymptotic flat spacetime in linear order, our approach is valid for GWs with any wavelengths in any order of metric perturbations.
As a byproduct, with our approach the gauge dependence issue on the energy spectrum of GWs disappears naturally~\cite{Cai:2021ndu}.
The reference spacetime for quasilocal energy is chosen to be Minkowski spacetime in this paper, a generalization to the de Sitter or anti-de Sitter spacetime is developed in~\cite{Chen:2016bgx}, further studies on this generalization will be done in the future.

\begin{acknowledgments}
Prof. Padmanabhan significantly contributed to a broad spectrum of topics related to astrophysics, cosmology, and classical and quantum aspects of gravitation.
It was a pity that Paddy passed away due to cardiac arrest.
It was a loss of our community.
As one of his friends, RGC would like to contribute this work to the Topical Collection (TC) ``In Memory of Prof. T. Padmanabhan''.
We thank Misao Sasaki and Shing-Tung Yau for helpful communications.
This work is supported in part by the National Natural Science Foundation of China Grants No.11690022, No.11821505, No.11991052, No.11947302 and by the Strategic Priority Research Program of the CAS Grant No.XDPB15, and by the Key Research Program of Frontier Sciences of CAS.
\end{acknowledgments}

\appendix

\section{Canonical form of Weyl tensor}\label{apdx:can_form_Weyl}

\begin{table}[htbp]
    \centering
    \caption{Canonical form of Weyl tensor for different Petrov types.
        The symbol $\not=$ means the corresponding Weyl scalar is nonvanishing.
    }
    \label{tab:can_form_Weyl}
    \begin{tabular}{|c|c|c|c|c|c|}
        \hline
    & $\Psi_{0}$ & $\Psi_{1}$ & $\Psi_{2}$ & $\Psi_{3}$ & $\Psi_{4}$ \\
        \hline
        I &$\not=$ &0 &$\not=$ &0 &$\not=$ \\
        \hline
        D &0 &0 &$\not=$ &0 &0 \\
        \hline
        II &0 &0 &$\not=$ &0 &$\not=$ \\
        \hline
        N &0 &0 &0 &0 &$\not=$ \\
        \hline
        III &0 &0 &0 &$\not=$ &0 \\
        \hline
        O &0 &0 &0 &0 &0 \\
        \hline
    \end{tabular}
\end{table}

\section{Tetrad transformations}\label{apdx:tetrad_trans}

The Lorentz transformation with six parameters acting on a null tetrad $\{m^{a},\br{m}^{a},l^{a},k^{a}\}$ can be classified to three types:

(i) $l'=l, \quad k'=k+a\br{m}+\br{a}m+a\br{a}l, \quad m'=m+al$

(ii) $k'=k, \quad l'=l+b\br{m}+\br{b}m+b\br{b}k, \quad m'=m+bk$

(iii) $k'=Ak, \quad l'=A^{-1}l, \quad m'=\ee^{\ii\theta}m$

\noindent where $a$ and $b$ are complex parameters, $A>0$ and $\theta$ are real parameters.

Data Availability Statement: Data sharing not applicable to this article as no datasets were generated or analysed during the current study.

\bibliography{citeLib}

\end{document}